\documentclass[12pt]{article}

\usepackage[iso-8859-7]{inputenc}
\usepackage{epsfig}
\usepackage{graphicx}
\usepackage{epstopdf}
\usepackage{amsfonts}
\usepackage{amssymb}
\usepackage{amsthm}
\usepackage{amsmath}
\usepackage{multirow}

\usepackage{color}

\newcommand{\newc}{\newcommand}
\newc{\ra}{\rightarrow}
\newc{\lra}{\leftrightarrow}
\newc{\be}{\begin{equation}}
\newc{\ee}{\end{equation}}
\newc{\bs}{\begin{split}}
\newc{\es}{\end{split}}
\newc{\ba}{\begin{eqnarray}}
\newc{\ea}{\end{eqnarray}}
\newc{\ov}{\overline}
\newc{\pa}{\partial}
\newc{\D}{\Delta}

\newc{\nn}{\nonumber}
\begin{document}
	\begin{titlepage}
		
		%\vspace*{-15mm}
		%\begin{flushright}
		%SHEP-11-XX\\
		%\end{flushright}
		\vspace*{0.7cm}

		\begin{center}
			{\large	\bf Non-Minimal Quartic Inflation in Supersymmetric $SO(10)$  }
			\\[12mm]
			George~K.~Leontaris$^{a,b}$
					\footnote{E-mail: \texttt{leonta@uoi.gr}},
		Nobuchika Okada$^{c}$
					\footnote{E-mail: \texttt{okadan@ua.edu}}
						and	Qaisar~Shafi$^{d}$
									\footnote{E-mail: \texttt{shafi@bartol.udel.edu}
						}
			\\[-2mm]
			
		\end{center}
		\vspace*{0.50cm}
			\centerline{$^{a}$ \it
					Theory Department, CERN,}
				\centerline{\it
					CH-1211, Geneva 23, Switzerland }
				\vspace*{0.15cm}
		\centerline{$^{b}$\it Physics Department, Theory Division, Ioannina University,}
				\centerline{\it
					GR-45110 Ioannina, Greece	}
		\vspace*{0.2cm}
			\centerline{$^{c}$ \it Department of Physics and Astronomy, University of Alabama,}
							\centerline{\it
	Tuscaloosa, Alabama 35487, USA	}
				\vspace*{0.15cm}
					\centerline{$^{d}$ \it 	Bartol Research Institute, Department of Physics and Astronomy, University of Delaware,}
							\centerline{\it
								DE 19716,  Newark, USA}
		\vspace*{1.00cm}
		
		\begin{abstract}
			\noindent
	
We describe how quartic ($\lambda \phi^4$) inflation with non-minimal coupling to gravity is realized in realistic supersymmetric 
$SO(10)$ models. In a well-motivated example the $16-\overline{16}$ Higgs multiplets, which break  $SO(10)$ to $SU(5)$ and yield 
masses for the right-handed neutrinos, provide the inflaton field $\phi$. Thus, leptogenesis is a natural outcome  in this class 
of $SO(10)$ models. Moreover, the adjoint (45-plet) Higgs also acquires a GUT scale value during inflation  so that the monopole 
problem is evaded. The scalar spectral index $n_s$ is in good agreement with the observations and $r$, the tensor to scalar ratio,
 is predicted for realistic values of GUT parameters to be of order $10^{-3}-10^{-2}$.
 % GUT parameters to lie close to 0.003.
  
\end{abstract}
		
	\end{titlepage}
	
	\thispagestyle{empty}
	\vfill
	\newpage

	%%%%%%%%%%%%%%%%%%    

%\section{Introduction}

By incorporating a single right-handed neutrino per generation to cancel new anomalies from gauging the accidental 
global U(1)$_{B-L}$ symmetry of the Standard Model (SM), both $SU(4)\times SU(2)_L\times SU(2)_R$~\cite{Pati:1974yy} 
and $SO(10)$~\cite{so10} provide particularly compelling examples of unifying the strong and electroweak forces. A 
non-supersymmetric model of SO(10) inflation~\cite{Lazarides:1991wu}, based on an earlier $SU(5)$ model~\cite{Shafi:1983bd}, 
was proposed a longtime ago.  In this class of $SO(10)$ inflation models, driven by a gauge  singlet field with minimal 
coupling to gravity and utilizing the Coleman-Weinberg potential~\cite{Coleman:1973jx}, the scalar to tensor ratio $r$, 
a canonical measure  of gravity waves generated during inflation, is estimated to be $\gtrsim$ 0.02, for 
$n_s= 0.96-0.97$~\cite{Shafi:2006cs}.  Depending on the $SO(10)$ symmetry breaking pattern, an observable number density 
of intermediate mass magnetic monopoles may be present in our galaxy~\cite{Lazarides:1984pq}.

In this letter  we propose to implement primordial inflation in realistic supersymmetric $SO(10)$ models~\cite{Pallis:2012zd}.  
We do this with a supergravity generalization of non-minimal $\lambda \phi^4$ inflation~\cite{Okada:2010jf}. Recall that 
$\lambda \phi^4$ inflation with a minimal coupling  to gravity predicts an $r$ value close to $0.25-0.3$, depending on the 
number of e-foldings ($N_0=60-50$). This prediction for $r$ lies well outside the 2-$\sigma$ range allowed by Planck~\cite{Ade:2015xua}
 and WMAP 9~\cite{Hinshaw:2012aka}. In contrast, $\lambda\phi^4$ inflation with a suitable non-minimal coupling to gravity 
 is in good agreement with the data regarding the key parameters $n_s$ and $r$. The quantity $r$, in particular, can be as 
 low as 0.003 or so, for $n_s = 0.96-0.97$. The discussion closely follows a previous model~\cite{Arai:2011nq} based on 
 supersymmetric $SU(5)$.

In order to retain perturbative unification of the MSSM gauge couplings in supersymmetric $SO(10)$  we prefer to work with
 lower dimensional  $SO(10)$ representations. We employ $16-\overline{16}$ Higgs to break $SO(10)$ to $SU(5)$ while keeping supersymmetry 
unbroken. The $\overline{16}$ vacuum expectation value (VEV) also provides large masses ($\lesssim 10^{14}$ GeV), via higher 
dimensional operators, to the right-handed neutrinos. In addition, the adjoint 45-plet, in conjunction either with a 54-plet or 
using higher dimensional operators, is employed to complete the breaking of $SO(10)$ to the MSSM gauge symmetry.  
Finally, following~\cite{Chang:2004pb}, we can employ two Higgs 10-plets to implement electroweak symmetry breaking and accommodate 
the charged fermion masses and mixings as well as neutrino oscillation data.  
This summarizes the basic structure of a realistic supersymmetric $SO(10)$ model.

Recall that a non-minimal $\lambda \phi^4$ inflation scenario is defined by the following action in the Jordan frame:
\begin{eqnarray}
{\cal S}_J= \int d^4 x\sqrt{-g}\left[-\frac 12(1+\xi\varphi^2) \, {\cal R}+
\frac 12g^{\mu\nu}\partial_{\mu}\varphi\partial_{\nu}\varphi- \frac{\lambda}{16} \varphi^4\right]\, , 
\label{NMI}
\end{eqnarray}
where we have set to unity  the reduced Planck mass, $M_P=2.44 \times 10^{18}$ GeV. In the limit $\xi\to 0_+$ the 
non-minimal gravitational coupling term $\xi\varphi^2 R$ vanishes and we approach minimal $\lambda \varphi^4$ chaotic 
inflation. In the Einstein frame with a canonical gravity sector, we can describe the action with a new inflaton field 
($\sigma$) which has a canonical kinetic term. The relation between $\sigma $ and $\varphi$ is given by
\begin{eqnarray}
\left(\frac{d\sigma}{d\varphi}\right)^{2} =
 \frac{1+ \xi (6 \xi +1) \varphi^2} {\left( 1 + \xi \varphi^2 \right)^2} \; .
\end{eqnarray}
The action in the Einstein frame is then given by
\begin{eqnarray}
S_E = \int d^4 x \sqrt{-g_E}\left[-\frac12  {\cal R}_E+\frac12 (\partial \sigma)^2
-V_E(\sigma(\varphi)) \right],
\end{eqnarray}
with 
\begin{eqnarray}
V_E =  \frac{\lambda}{16} \frac{\varphi^4}{(1+\xi \varphi^2)^2}. 
\label{VE}
\end{eqnarray}

The slow-roll parameters in terms of the original scalar field ($\varphi$) are expressed as 
\begin{eqnarray}
\epsilon(\varphi)&=&\frac{1}{2} \left(\frac{V_E'}{V_E \sigma'}\right)^2, 
 \nonumber \\
\eta(\varphi)&=& 
\frac{V_E''}{V_E\, (\sigma')^2}- \frac{V_E'\sigma''}{V_E\, (\sigma')^3} ,  
 \nonumber \\
\zeta (\varphi) &=&  \left(\frac{V_E'}{V_E \sigma'}\right) 
 \left( \frac{V_E'''}{V_E\, (\sigma')^3}
-3 \frac{V_E''\, \sigma''}{V_E\, (\sigma')^4} 
+ 3 \frac{V_E'\, (\sigma'')^2}{V_E\, (\sigma')^5} 
- \frac{V_E'\, \sigma'''}{V_E\, (\sigma')^4} \right)  , 
\end{eqnarray}
where a prime denotes a derivative with respect to $\varphi$. The amplitude of the curvature perturbation 
$\Delta_{\mathcal{R}}$ is given by 
\begin{equation} 
\Delta_{\mathcal{R}}^2 = \left. \frac{V_E}{24 \pi^2 \epsilon } \right|_{k_0},
\end{equation}
 with $\Delta_\mathcal{R}^2= 2.195 \times10^{-9}$ from the Planck measurement~\cite{Ade:2015xua} with the pivot 
 scale chosen at $k_0 = 0.002$ Mpc$^{-1}$. The number of e-folds is given by
\begin{eqnarray}
  N_0 = \frac{1}{\sqrt{2}} \int_{\varphi_{\rm e}}^{\varphi_0}
\frac{d\varphi}{\sqrt{\epsilon(\varphi)}}\left(\frac{d\sigma}{d\varphi}\right)\,,
\end{eqnarray} 
where $\varphi_0$ is the inflaton value at horizon exit of the scale corresponding to $k_0$, and $\varphi_e$ is the 
inflaton value at the end of inflation, which is defined by ${\rm max}[\epsilon(\varphi_e), | \eta(\varphi_e)| ]=1$.
The value of $N_0$ depends logarithmically on the energy scale during inflation as well as on the reheating temperature, 
and is typically taken to be $N_0=50-60$.

%%%%%%%%%%%%%%%%%%%%%%%%%%%%%%%%%%%%%%%%%%%%
\begin{table}[ht]
\begin{center}
\begin{tabular}{|c||cc|ccc|c|}
\hline
\multicolumn{7}{|c|}{$N_0=60$}   \\
\hline 
 $\xi $ &  $\varphi_0$ & $\varphi_e $ & $n_s$   &  $r$  &  $-\alpha (10^{-4})$  & $\lambda$\\ 
\hline
  $0$         & $22.2$   & $3.46$    &  $0.951$ &   $ 0.260$    & $-7.93 $    &  $ 5.59 \times 10^{-13}  $            \\
  $0.001$  & $22.2$   & $3.43$    & $0.957$  &  $ 0.174$     & $-7.650$  &   $ 8.36  \times 10^{-13}  $           \\
  $0.01$    & $21.7$   & $3.18$    & $0.965$  &  $ 0.0451$   & $-6.12$    &   $ 3.45  \times 10^{-12}  $           \\
  $0.1$      & $17.8$   & $2.15$    & $0.967$  &  $ 0.00784$ & $-5.39$    &   $ 4.34  \times 10^{-11}  $            \\
  $1$         & $8.52$   & $1.00$    & $0.968$  &  $0.00346$  & $-5.25$    &   $ 1.85  \times 10^{-9}  $             \\
  $10$       & $2.89$   & $0.337$  & $0.968$  &  $0.00301$  & $-5.24$    &   $ 1.60  \times 10^{-7}  $              \\
 $100$      & $0.920$ & $0.107$  & $0.968$  &  $0.00297$  & $-5.23$    &   $ 1.58  \times 10^{-5}  $              \\
$252$      & $0.580$ & $0.0677$  & $0.968$  &  $0.00297$  & $-5.23$    &   $ 1.0  \times 10^{-4}  $              \\
$1000$     & $0.291$ & $0.0340$ & $0.968$  &  $0.00296$  & $-5.23$    &   $ 1.58  \times 10^{-3}  $              \\
$10000$   & $0.0921$ & $0.0107$ & $0.968$  &  $0.00296$  & $-5.23$    &   $ 0.158    $              \\
\hline
\end{tabular}
\end{center}
\caption{ 
Inflationary predictions for various $\xi$ values in $\lambda \phi^4$ inflation with non-minimal gravitational coupling. 
} 
\label{Tab:1}
\end{table}
%%%%%%%%%%%%%%%%%%%%%%%%%%%%%%%%%%%%%%%%%%%%%%%%%%%%%%

The slow-roll approximation is valid as long as the conditions $\epsilon \ll 1$, $|\eta|\ll 1$ and $\zeta\ll 1$ hold. 
In this case, the scalar spectral index $n_{s}$, the tensor-to-scalar ratio $r$, and the running of the spectral index 
$\alpha=\frac{d n_{s}}{d \ln k}$, are given by
\begin{eqnarray}
n_s = 1-6\epsilon+2\eta, \; \; 
r = 16 \epsilon,  \; \;
\alpha=16 \epsilon \eta - 24 \epsilon^2 - 2 \zeta. 
\end{eqnarray} 
Here the inflationary predictions are evaluated at $\varphi=\varphi_0$. With the constraint $\Delta_\mathcal{R}^2= 2.215
\times10^{-9}$, once $N_0$ is fixed, the inflationary predictions as well as the quartic coupling $\lambda$ are determined 
as a function of $\xi$. In Table~\ref{Tab:1} we list the numerical results for selected values of $\xi$. The inflationary 
predictions are consistent with the Planck results  ($n_s=0.9655 \pm 0.0062$, $r \lesssim 0.07$ and $\alpha=0.0057\pm 0.0071$
at 68\% C.L.) for $\xi \gtrsim 0.01$. As $\xi$ increases, the inflationary predictions approach $n_s \simeq 0.968$,  $r 
\simeq 0.00296$ and $\alpha \simeq -5.23 \times 10^{-4}$,  while the quartic coupling is monotonically increasing.

Next we discuss how this scenario is implemented in a realistic supersymmetric $SO(10)$ model. The relevant superpotential 
terms for inflation are given by
\be 
{\cal  W}\supset  \frac 12 m_A A^2 +\bar z(m- y A)z-\frac 12 m_A\left(\frac{m}{y}\right)^2\,,
\ee 
where $z, \bar z$ denote the  $16-\overline{16}$ fields, $A$ represents the 45-plet, and the last term has  been included 
so that $\langle {\cal  W}\rangle  = 0 $ at the desired supersymmetric minimum with $SO(10)$ broken to the SM gauge group. 
To implement non-minimal $\lambda \phi^4$ inflation an appropriate K\"ahler  potential, following \cite{Ferrara:2010yw}, 
is given by  
\be 
\Phi =1-\frac 13 \left( |\bar z|^2+ |z|^2+|A|^2 \right) +\frac 13 \gamma (\bar z z+h.c.)+\frac 16 \gamma_A(A^2+h.c.) \,,
\ee 
where the parameter coefficients $\gamma$ and $\gamma _A$  are assumed to be real and positive constants. 

The inflaton trajectory is parametrized by the $D$-flat direction
\be
   \bar z= z ={\scriptsize \frac 12}\varphi, \;  A=\frac{a}{\sqrt{2}} 
   \label{fdzzbA}\,,
\ee
where the field  VEVs  $\varphi$ and $a$  break $SO(10) \to SU(5)$ and  $SO(10) \to SU(3) \times SU(2)_L \times 
SU(2)_R \times U(1)_{B-L}$, respectively. Thus, the initial theory reduces to a model with two real scalars 
$\varphi$ and $a$, and the Jordan frame action is given by 
\begin{eqnarray}
{\cal S}= \int d^4 x\sqrt{-g} \left[ 
-\frac 12  \Phi {\cal R}+
\frac 12g^{\mu\nu} (\partial_{\mu}\varphi)(\partial_{\nu}\varphi) +
\frac 12g^{\mu\nu} (\partial_{\mu}a)(\partial_{\nu} a) -V_J  \right]  \, .
\end{eqnarray}
Here, the K\"ahler potential is expressed as  
\begin{eqnarray}
 \Phi = 1 + \xi  \varphi^2 + \xi_A a^2 , 
\end{eqnarray}
where $\xi =(\gamma -1)/6$ and $\xi_A=(\gamma_A -1)/6$. The scalar potential $V_J$ in the Jordan frame is calculated 
as~\cite{SUGRA}
\begin{eqnarray}
   V_J  &=& 
 - \left(
\begin{array}{cccc}
3 {\cal W}  &  \frac{\partial\cal W}{\partial \bar{z}} &  \frac{\partial\cal W}{\partial z}
   &  \frac{\partial\cal W}{\partial A}
\end{array}
\right) {\cal M}^{-1}
 \left(
\begin{array}{cccc}
3 {\cal W}  &  \frac{\partial\cal W}{\partial \bar{z}} &  \frac{\partial\cal W}{\partial z}
   &  \frac{\partial\cal W}{\partial A}
\end{array}
\right)^\dagger \; , \label{VJ}
\end{eqnarray}   
where ${\cal M}^{-1}$ is the inverse of the matrix 
\begin{eqnarray}
{\cal M}=3 \left(
\begin{array}{cccc}
\Phi  & \frac{\partial \Phi}{\partial {\bar z}}  &  \frac{\partial \Phi}{\partial z} & \frac{\partial \Phi}{\partial A}    \\
\frac{\partial \Phi}{\partial {\bar z}^\dagger} & 
\frac{\partial^2 \Phi}{\partial {\bar z}^\dagger \partial {\bar z}}  &
\frac{\partial^2 \Phi}{\partial {\bar z}^\dagger \partial z}   &
\frac{\partial^2 \Phi}{\partial {\bar z}^\dagger \partial A}     \\
\frac{\partial \Phi}{\partial z^\dagger} & 
\frac{\partial^2 \Phi}{\partial z^\dagger \partial {\bar z}}  &
\frac{\partial^2 \Phi}{\partial z^\dagger \partial z}   &
\frac{\partial^2 \Phi}{\partial z^\dagger \partial A}     \\
\frac{\partial \Phi}{\partial A^\dagger} & 
\frac{\partial^2 \Phi}{\partial A^\dagger \partial {\bar z}}  &
\frac{\partial^2 \Phi}{\partial A^\dagger \partial z}   &
\frac{\partial^2 \Phi}{\partial A^\dagger \partial A}     
\end{array}
\right).  
\end{eqnarray}
To compute the potential~(\ref{VJ}) we write ${\cal W} $ in terms of $\varphi$ and $a$ 
\begin{eqnarray} 
{\cal W}  &=& \frac{1}{4} m_A  \left( a^2 - 2 \left( \frac{m}{y} \right)^2 \right) +
                \frac{1}{4} \varphi^2 \left( m - \frac{y}{\sqrt{2}} a \right) \,.    \nonumber
                \end{eqnarray} 
Then  we have 
\begin{eqnarray}
\frac{\partial\cal W}{\partial \bar{z}} &=&  \frac{\partial\cal W}{\partial z} = 
\frac{1}{2} \varphi  \left( m - \frac{y}{\sqrt{2}} a \right),   \nonumber \\ 
\frac{\partial\cal W}{\partial A} &=& \frac{m_A}{\sqrt{2}} a - \frac{y}{4} \varphi^2 ,  \nonumber 
\end{eqnarray} 
and
\begin{eqnarray}
{\cal M} &=& 3 \left(
\begin{array}{cccc}
 \Phi  &   \xi \varphi  &    \xi \varphi  &  \sqrt{2} \xi_A a   \\
  \xi \varphi  &  -1/3  &  0  &  0   \\ 
  \xi \varphi  &  0  &  -1/3  &  0   \\ 
 \sqrt{2} \xi_A a  & 0 & 0 & -1/3 \\
\end{array}
\right).  
\end{eqnarray}   
The potential minimum where $SO(10)$ is broken to the SM gauge group lies at 
\begin{eqnarray}    
    \phi=2\frac{\sqrt{m \; m_A}}{y}, \; \; \;  a=\sqrt{2} \; \frac{m}{y} .
\label{vacuum}
\end{eqnarray}

%%%%%%%%%%%%%%%%%%%%%%%%%%%%%%%%%%%%%%%%%%%%
%   Fig 1
%%%%%%%%%%%%%%%%%%%%%%%%%%%%%%%%%%%%%%%%%%%%
\begin{figure}[t]
\begin{center}
\includegraphics[width=0.465\textwidth,angle=0,scale=1.05]{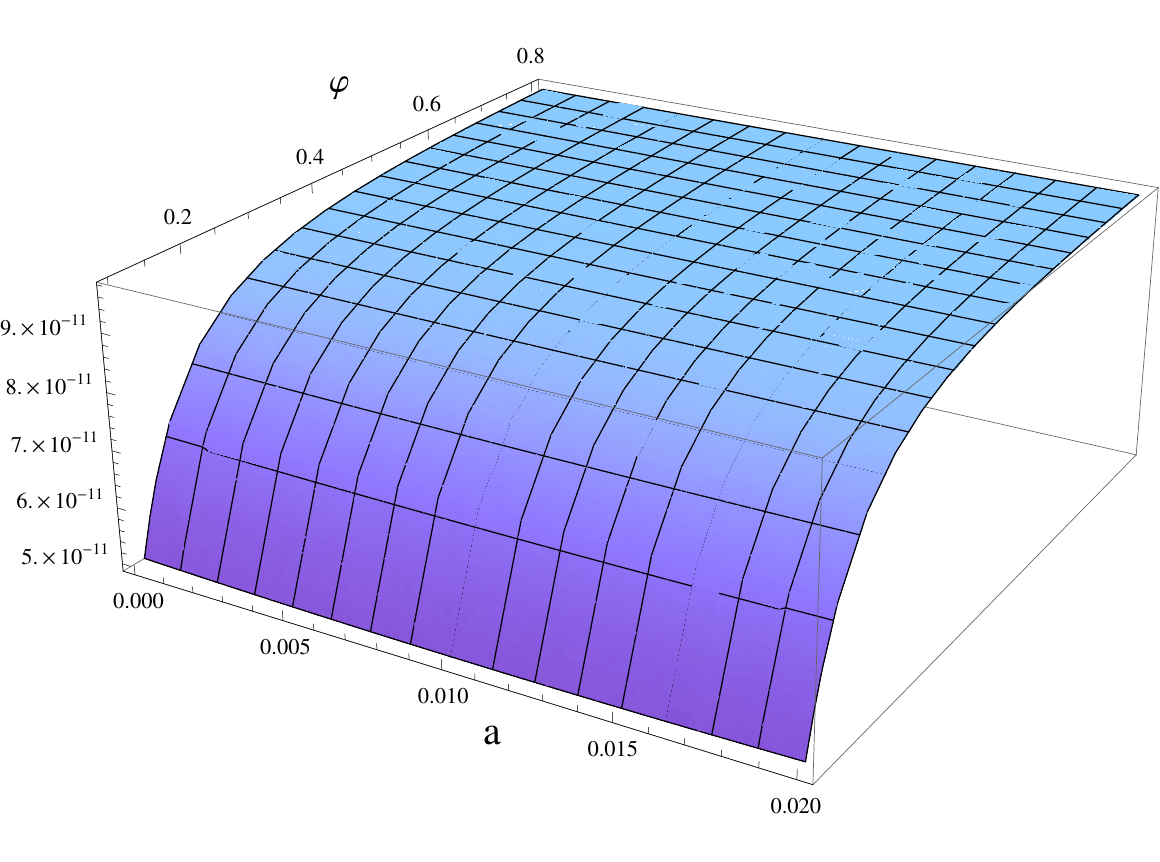}
\hspace{0.1cm}
\includegraphics[width=0.465\textwidth,angle=0,scale=1.05]{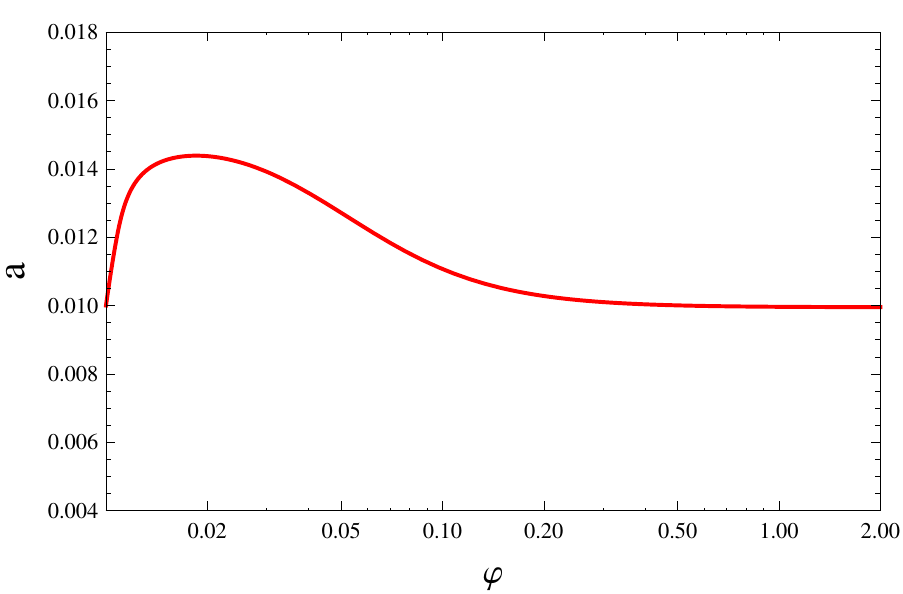}
\end{center}
\caption{
The scalar potential $V_E$ in the Einstein frame (left) and the inflaton trajectory (right). 
Here, we have fixed the parameters as $\xi=252$, $\xi_A=-68$, $m=\frac{y}{\sqrt{2}} M_G$ 
   and $m_A=\frac{y}{2 \sqrt{2}} M_G$ { with $y=0.01$} and $M_G=0.01 $ (typical GUT scale).   
}
\label{fig:1}
\end{figure}
%%%%%%%%%%%%%%%%%%%%%%%%%%%%%%%%%%%%%%%%%%%% 

The dynamics of inflation is encoded in the scalar potential $V_E = V_J/\Phi^2$ in the Einstein frame.  
In Fig.~\ref{fig:1} we show a 3-dimensional plot of $V_E$ (left panel) and  the inflaton trajectory (right 
panel). Here, we have fixed the parameters as $\xi=252$, $\xi_A=-68$, $m=\frac{y}{\sqrt{2}} M_G$ and $m_A=
\frac{y}{2 \sqrt{2}} M_G$    with, $y=0.01$ and $M_G=0.01$  a typical value for the GUT scale. The right 
panel indicates that for $\varphi \gtrsim 0.1$ the inflaton trajectory is well approximated as a straight 
line such that the $\varphi$ field is identified with the inflaton. Note that along this  trajectory for 
$\varphi \gtrsim 0.1$,  $a$ stays nearly constant close to its value at the potential minimum, $a=M_G$ 
(see Eq.~(\ref{vacuum})). In this case, the scalar potential along the trajectory is greatly simplified as 
\begin{eqnarray}
V_E &\simeq & \frac{y^2}{16}\left( \frac{\varphi^2  - M_G^2}{1+\xi \varphi^2+\xi_A M_G^2 } \right)^2 
\frac{1+\xi (6 \xi+1) \varphi^2 + \xi_A M_G^2}{1+\xi (6 \xi+1) \varphi^2 + \xi_A (6 \xi_A+1) M_G^2} \nonumber \\
& \simeq & \frac{y^2}{16}  \frac{\varphi^4}{\left( 1+\xi \varphi^2  \right)^2 }. 
\end{eqnarray}
Here we have used $\xi > \xi_A$ and $\varphi^2 \gg M_G^2$ for $\varphi \gtrsim 0.1$. This potential is exactly the 
same as Eq.~(\ref{VE}) with the identification  $\lambda=y^2$. Since the inflaton value at the end of inflation is 
found to be $\varphi_e=0.677$ for $\xi=252$ (see Table~\ref{Tab:1}), the displacement of $a$ during inflation is small 
and hence our inflation scenario in the context of supergravity  is well approximated by  $\lambda \phi^4$ inflation 
with non-minimal gravitational coupling. Table~\ref{Tab:1} shows the inflationary predictions as  $n_s \simeq 0.968$, 
 $r \simeq 0.00297$ and $\alpha \simeq -5.23 \times 10^{-4}$, which are consistent with the Planck results. 
Along the inflaton trajectory $SO(10)$ is broken to the SM, and hence the primordial monopoles are inflated away.

In our analysis, we have set $y=0.01$. We find that the shape of the inflaton trajectory shown in the right panel of 
Fig.~\ref{fig:1} is almost unchanged for a variety of choices of the model parameters, $y, \xi$ and $\xi_A$.~\footnote{We 
find that $\xi_A$ must be negative in order to bound the scalar potential from below in the $a$-direction.} In order to 
identify $\varphi$ with the inflaton, the condition $\varphi_e \gtrsim 0.1$ is crucial. According to the results listed in 
Table~\ref{Tab:1}, this  means $\xi \lesssim 100$, or equivalently $\lambda = y^2 \lesssim 10^{-4}$.  Following the $SO(10)$ 
symmetry breaking to the SM, the components $10 + \overline{10}$ of $SU(5)$ from $16, \overline{16}$ and 45 fields, have 
masses of  ${\cal O}(y M_G) \lesssim 10^{14}$ GeV. There is some mass splitting of the same order within these multiplets 
but gauge coupling unification is essentially preserved.  With the intermediate scale $y M_G$ of order $10^{11} - 10^{14}$  
GeV, the tensor to scalar ratio $r$ varies between  0.01 to 0.003 which should be testable in the foreseeable future.

The VEV of the $\overline{16}$ Higgs field not only breaks the $SO(10)$ symmetry but also generates  Majorana masses 
for the right-handed neutrinos through  higher dimensional operators of the form, 
\begin{eqnarray}
 {\cal W} \supset \frac{c_i}{M_P} {\bf 16}_i {\bf 16}_i  \bar{z} \bar{z} ,
\end{eqnarray}
where ${\bf 16}_i $ $(i=1, 2, 3)$ denotes the matter field, and the coefficient $c_i$ is taken to be flavor-diagonal. 
Associated with the $SO(10)$ symmetry breaking, the right-handed neutrinos acquire masses $M_i = c_i M_G^2/M_P = c_i 
m_\varphi \simeq c_i \times 10^{14}$ GeV,  where $m_\varphi=y M_G =M_G^2/M_P$  is the inflaton mass. 

Another important role of the higher dimensional operators is that after inflation the inflaton $\varphi$ decays 
into right-handed neutrinos through them to reheat the Universe. We estimate the reheating temperature as 
\begin{eqnarray}
  T_{RH} \simeq  \sqrt{\Gamma_\varphi M_P} \simeq \frac{1}{\sqrt{16 \pi}} M_3 = \frac{|c_3|}{\sqrt{16 \pi}} m_\varphi,  
\end{eqnarray}
where $M_3$ is the heaviest right-handed neutrino mass, compatible with kinematics, and 
\begin{eqnarray}
   \Gamma_\varphi \simeq \frac{1}{16 \pi}  \left( \frac{M_3}{M_G}\right)^2 m_\varphi 
\end{eqnarray} 
is the total decay width of the inflaton (assuming $c_3 <1/2$). Since $T_{RH} < M_3$, we expect that the reheating occurs 
after scatterings among the produced heavy neutrinos and their decays, and hence the actual reheating temperature is lower 
than the value estimated above. In order to avoid the cosmological gravitino problem~\cite{GravitinoProb}, we consider the 
upper bound on the reheating temperature of $T_{RH} < 10^6-10^9$ GeV with the gravitino mass in the range of $100$ GeV$\lesssim 
m_{\tilde{G}} \lesssim 10$ TeV~\cite{Kawasaki:2008qe}, and take $c_3$ small enough to satisfy this upper bound.  Depending on the 
value of reheating temperature and the right-handed neutrino mass spectrum we can consider either thermal~\cite{Fukugita:1986hr}
or non-thermal~\cite{Lazarides:1991wu} letogenesis scenarios.

In summary, we have shown that $\lambda \phi^4$ inflation with non-minimal coupling to gravity  can be realized in the framework 
of realistic supersymmetric $SO(10)$ models. An attractive feature is the utilization as inflaton of a field already present for 
particle physics reasons. In the example provided inflation is driven by the field that breaks $SO(10)$ to $SU(5)$ and provides
 masses  to the right-handed neutrinos. Depending on additional details, thermal or non-thermal leptogenesis is a natural outcome. 
The field associated with monopole production is non-zero during inflation and  so these topological defects are inflated away. 
With a scalar spectral index in the vicinity of $0.96-0.97$ the tensor to scalar ratio $r$ is  estimated to be of order 
$10^{-3}-10^{-2}$.   Significantly larger values of $r$ require appreciably smaller values of the quartic coupling. 

\vspace{1cm}

\vspace{1cm} 
{\bf Acknowledgements} \\
{\it Q.S. thanks Steve Barr for helpful discussions.
G.K.L. would like to thank the Physics and Astronomy Department and Bartol Research Institute of the University of Delaware
 for kind hospitality. This work is supported in part by the DOE Grant Nos.~ DE-SC0013680 (N.O.) and DE-SC0013880 (Q.S.).}

\end{document}